




\headline={\ifnum\pageno=1\firstheadline\else
\ifodd\pageno\rightheadline \else\leftheadline\fi\fi}
\def\firstheadline{\hfil}
\def\rightheadline{\hfil}
\def\leftheadline{\hfil}
        \footline={\ifnum\pageno=1\firstfootline\else\otherfootline\fi}
\def\firstfootline{\rm\hss\folio\hss}
\def\otherfootline{\hfil}

\font\twelvebf=cmbx10 scaled\magstep 1
\font\twelverm=cmr10 scaled\magstep 1
\font\twelveit=cmti10 scaled\magstep 1


\font\tenrm=cmr10
\font\tenit=cmti10


\def\IiI{{\hbox{\tenrm I\kern-.19em{I}}}}

\def\uq2{U_q({\uit su}(2))}


\def\ggg{${\cal G} $}

\def\half{{1\over{2}}}

\def\CcC{{\hbox{\tenrm C\kern-.45em{\vrule height.67em width0.08em depth-.04em
\hskip.45em }}}}
\def\RrR{{\hbox{\tenrm I\kern-.17em{R}}}}
\def\HhH{{\hbox{\tenrm {I\kern-.18em{H}}\kern-.18em{I}}}}
\def\DdD{{\hbox{\tenrm {I\kern-.18em{D}}\kern-.36em {\vrule height.62em
width0.08em depth-.04em\hskip.36em}}}}

\def\ZzZ{{\hbox{\tenrm Z\kern-.31em{Z}}}}

\def\IiI{{\hbox{\tenrm I\kern-.19em{I}}}}
\def\NnN{{\hbox{\tenrm {I\kern-.18em{N}}\kern-.18em{I}}}}
\def\QqQ{{\hbox{\tenrm {{Q\kern-.54em{\vrule height.61em width0.05em
depth-.04em}\hskip.54em}\kern-.34em{\vrule height.59em width0.05em
depth-.04em}}
\hskip.34em}}}
\def\OoO{{\hbox{\tenrm {{O\kern-.54em{\vrule height.61em width0.05em
depth-.04em}\hskip.54em}\kern-.34em{\vrule height.59em width0.05em
depth-.04em}}
\hskip.34em}}}

\def\uq2{U_q({\uit su}(2))}

\def\fraz#1#2{{\strut\displaystyle #1\over\displaystyle #2}}

\def\part#1{\fraz{\partial}{\partial#1}}

\def\su2q{SU(2)_q}
\def\h1q{H(1)_q}

\def\nu{N_{1}}






\parindent=1.5pc
\hsize=6.0truein
\vsize=8.5truein
\nopagenumbers

\centerline{\bf  QUANTUM DISSIPATION AND COHERENCE}



\vglue 0.8cm

\centerline{\tenrm GIUSEPPE VITIELLO}
\baselineskip=13pt
\centerline{\tenit  Dipartimento di Fisica, Universit\`a di Salerno and
INFN Napoli,}
\baselineskip= 12 pt
\centerline{\tenit I84100 Salerno, Italy}
\baselineskip= 12 pt
\centerline{\tenit Vitiello@sa.infn.it}
\vglue 0.8cm
\centerline{\tenrm ABSTRACT}
\vglue 0.3cm
{\rightskip=3pc
 \leftskip=3pc
 \tenrm\baselineskip=12pt\noindent
We discuss
dissipative systems in Quantum Field Theory by studying
the canonical quantization of
the damped harmonic oscillator (dho). We
show that the
set of states of the system splits into unitarily inequivalent
representations of the canonical commutation relations.
The irreversibility of time evolution is expressed as tunneling among
the unitarily inequivalent representations.
Canonical quantization is shown to lead to time dependent SU(1,1)
coherent states. We derive the exact action for the dho
from the path integral formulation of the quantum Brownian motion
developed by Schwinger and by Feynman and Vernon. The doubling of
the phase-space degrees of freedom for dissipative systems is
related to quantum noise effects.
Finally, we express the time evolution
generator of the dho in terms of operators of the $q$-deformation of the
Weyl-Heisenberg algebra.
The $q$-parameter acts as a label for the unitarily inequivalent
representations.
\vglue 0.6cm}

\vfil
\twelverm
\baselineskip=14pt


\bigskip

\noindent {\twelvebf 1. Introduction}
\bigskip
\indent In this paper we want to report on
some recent work$^{1-5}$ on dissipative
systems in quantum theory.

In principle there is no room for dissipative
systems in Quantum Mechanics(QM) since the formalism of QM is
based on conserving
probabilities, and one has to introduce some sort of generalized
quantum formalism to describe damped systems.
The
developments of the theory of metastable states going beyond
the Breit-Wigner treatment and other phenomenological approachs
have been recently reported in ref. 6 where the generalized quantum
theory for unstable systems
of Sudarshan et al.$^{6,7}$ has been also reviewed.

Dissipative systems from the point of view of the quantum theory for
Brownian motion have been
analyzed in the path integral formalism by Schwinger$^{8}$
and by Feynman and Vernon$^{9}$
and are of course a major topic in nonequilibrium
statistical mechanics and nonequilibrium Quantum Field Theory (QFT) at finite
temperature$^{10,11}$.

The microscopic theory
for a dissipative system
must include the details of processes responsible for dissipation,
including quantum
effects. One may start since the beginning with a Hamiltonian that
describes the system, the bath and the system-bath interaction.
Subsequently, the description of the original dissipative system is
recovered by the reduced density matrix obtained by eliminating
the bath variables which originate the damping and the fluctuations.
The problem with dissipative systems in QM is indeed that
canonical commutation relations (ccr) are not preserved by time evolution
due to damping terms. The r\^ole of fluctuating forces is in fact the one
of preserving the canonical structure.

At a classical level, it is known since long time$^{12}$ that
the attempt to derive, from a variational
principle, the equations of motion defining the dissipative system
requires the introduction of additional complementary
equations.

This latter approach has been pursued in refs. 13 and 1-5
where the quantization
of the damped harmonic osscillator (dho) has been studied by doubling the
phase-space degrees of freedom.
The doubled degrees of freedom play the r\^ole of
the bath degrees of freedom. In sec. 2 we present such an approach.

We have shown in refs. 1-3 that
the dynamical
group structure associated with the canonical quantization of dho
is that of SU(1,1) and that time evolution would
lead out of the Hilbert space of states; in other words, the
quantum mechanical treatment of dho does not provide a unitary
irreducible representation of SU(1,1)$^{14}$. To
cure these pathologies one must move to QFT,
where infinitely many unitarily inequivalent ({\twelveit ui})
representations of the ccr are allowed (in the infinite volume or
thermodynamic limit). The reason for this is that
the set of the states of the damped oscillator splits
into {\twelveit ui}
representations (i.e. into disjoint {\twelveit folia},
in the C*-algebra formalism)
each one representing the states of the system at time {\twelveit t}:
in a more
conventional language, the time evolution may be described as
{\twelveit tunneling} between {\twelveit ui}
representations. A remarkable feature of our description thus emerges:
{\twelveit
at
microscopic level the irreversibility of time evolution (the
arrow of time) of
damped oscillator is expressed by the non unitary evolution across the
ui representations of the ccr.}

We stress that the nature of the ground states of the {\twelveit ui}
representations is the one of the  SU(1,1) coherent states. Furthermore, it
has been
shown$^{\, 1}$
that the squeezed coherent
states of light entering quantum optics$^{\, 15}$ can be identified, up
to elements of the group \ggg ~ of automorphisms of $su(1,1)$, with the
states of the quantum dho.

In ref. 3 it has been shown that the dho states are time dependent
thermal states, as expected due to the statistical nature of dissipation.
The formalism
for the dho turns out to be similar to the one of real time
QFT at finite temperature,
also called thermo-field dynamics (TFD)$^{11,16,17}$. In refs. 18 and
19 such
a connection with TFD has been further analysed and the master
equation has been discussed$^{18}$.

In ref. 4
the exact action for the dho in the path integral
formalism of Schwinger and Feynman and Vernon has been obtained.
In particular
the initial values of the doubled variables have been related
to the probability of quantum fluctuations in the ground state,
a result which is interesting also in the more general case
of thermal field theories. We report such results in sec. 3.

Finally,
in sec. 4,
we explicitly express$^{5}$ the
time evolution generator of
the dho
in terms of operators of
q-deformed Weyl-Heisenberg
(q-WH)$^{20}$
algebra.
The relation of q-WH
algebra with thermal field theory
is also commented upon.
The
q-deformation parameter turns out to be related with time
parameter in the
case of dho and  with temperature in the case of
thermal field theory. In both cases, the q-parameter acts
as a label for the {\twelveit ui} representations
of the ccr
in which the space of the states splits
in the infinite volume limit.
Such a conclusion confirms
a general analysis
$^{21}$ which shows that the Weyl representations
in QM and the {\twelveit ui} representations in
QFT are indeed labelled by the deformation parameter of
the q-WH algebra.
Sec. 5 is devoted to the conclusions.
\bigskip
\noindent {\twelvebf 2. The canonical quantization of the damped
harmonic oscillator}
\bigskip
We consider the damped harmonic oscillator with classical equation
$$
m \ddot x + \gamma \dot x + \kappa x = 0 \quad , \eqno{(2.1)}
$$
and we want to perform its canonical quantization.
We closely follow the approach of refs. 1-3 and 13.

In order to deal with an isolated system, as the canonical
quantization scheme requires, it is
necessary to double the the phase-space
dimensions$^{12,13}$. The lagrangian for system 2.1 is written as
$$
L = m \dot x \dot y + \half \gamma ( x \dot y - \dot x y ) - \kappa x y
\quad . \eqno{(2.2)}
$$

Eq.2.1 is obtained by varying eq. 2.2 with respect to $y$, whereas
variation with respect to $x$ gives
$$
m \ddot y - \gamma \dot y + \kappa y = 0 \quad , \eqno{(2.3)}
$$
\noindent which appears to be the {\twelveit time reversed} ($\gamma
\rightarrow - \gamma$) of eq. 2.1.
$y$ may be thought of as describing an
effective degree of freedom for the heat bath to which the system 2.1
is coupled.
The canonical momenta
are then given by ${\,
p_{x} \equiv {{\partial L}\over{\partial \dot x}} = m \dot y - \half
\gamma y}$ ; ${p_{y} \equiv {{\partial L}\over{\partial \dot y}}
= m \dot x + \half \gamma x}$.
The hamiltonian is
$$
H = p_{x} \dot x + p_{y} \dot y - L = {{1}\over{m}} p_{x} p_{y} +
{{1}\over{2m}}\gamma
\left ( y p_{y} - x p_{x} \right ) + \left ( \kappa - {{\gamma^{2}}\over{4 m}}
\right ) x y \quad .
\eqno{(2.4)}
$$
Canonical quantization is then performed by introducing the
commutators
$[ x , p_{x} ]=
i\, \hbar = [ y , p_{y} ] , ~
[ x , y ] = 0 = [ p_{x} , p_{y} ]$, and the corresponding sets of
annihilation and creation operators
$$
\eqalign{
\alpha &\equiv \left ({1\over{2 \hbar \Omega}} \right )^{1\over{2}} \left (
{{p_{x}}\over{\sqrt{m}}} - i \sqrt{m} \Omega x \right ) \quad , \quad
\alpha^{\dagger} \equiv \left ({1\over{2 \hbar \Omega}} \right )^{1\over{2}}
\left ( {{p_{x}}\over{\sqrt{m}}} + i \sqrt{m} \Omega x \right ) \quad , \cr
\beta &\equiv \left ({1\over{2 \hbar \Omega}} \right )^{1\over{2}} \left (
{{p_{y}}\over{\sqrt{m}}} - i \sqrt{m} \Omega y \right ) \quad , \quad
\beta^{\dagger} \equiv \left ({1\over{2 \hbar \Omega}} \right )^{1\over{2}}
\left ( {{p_{y}}\over{\sqrt{m}}} + i \sqrt{m} \Omega y \right ) \quad ,}
\eqno{(2.5a)}
$$

$$
[\, \alpha , \alpha^{\dagger} \,] = 1 = [\, \beta , \beta^{\dagger} \, ] \quad
, \quad  [\, \alpha , \beta \,] = 0 = [\, \alpha , \beta^{\dagger} \, ] \quad .
\eqno{(2.5b)}
$$
\noindent We have introduced ${\Omega \equiv \left [
{1\over{m}} \left ( \kappa - {{\gamma^{2}}\over{4 m}} \right )
\right ]^{1\over{2}}}$, the common frequency of the two oscillators eq. 2.1
and eq. 2.3, assuming $\Omega$ real, hence ${ \kappa >
{{\gamma^{2}}\over{4 m}}}$ (case of
no overdamping).
The Feshbach
and Tikochinsky $^{13}$ quantum hamiltonian is then obtained as
$$
H = \hbar\Omega(\alpha^{\dagger}\beta + \alpha \beta^{\dagger})
- {i\hbar\gamma\over{4m}}\left[(\alpha^{2} - \alpha^{\dagger 2})
 - (\beta^{2} - \beta^{\dagger 2})\right]  ~~.
\eqno{(2.6)}
$$

In sec. 3 we show that, at quantum level, the $\beta$ modes
allow quantum noise effects arising from the imaginary part
of the action$^{4}$.
By using the canonical linear transformations
$
{A \equiv {1\over{\sqrt 2}}
( \alpha + \beta )}
,$~
$
{B \equiv {1\over{\sqrt 2}}
( \alpha - \beta )},~
$
$H$ is written as
$$
\eqalign{
 H &=  H_{0} +  H_{I} \quad , \cr
 H_{0} = \hbar \Omega ( A^{\dagger} A - B^{\dagger} B ) \quad &, \quad
H_{I} = i \hbar \Gamma ( A^{\dagger} B^{\dagger} - A B ) \quad ,\cr}
\eqno{(2.7)}
$$
\noindent where the decay constant for the classical variable $x(t)$ is
denoted by $\Gamma \equiv {{\gamma}\over{2 m}}$.

We note that the states generated by $B^{\dagger}$ represent
the sink where the energy
dissipated by the quantum damped oscillator flows: the $B$-oscillator
represents the reservoir or heat bath coupled to the
$A$-oscillator.

The dynamical group structure associated
with the system of coupled quantum oscillators is that of $SU(1,1)$.
The
two mode realization of the algebra $su(1,1)$ is indeed generated by
$
J_{+} = A^{\dagger} B^{\dagger} \quad , \quad J_{-} = J_{+}^{\dagger} = A B
\quad , \quad
J_{3} = {1\over{2}} (A^{\dagger} A + B^{\dagger} B + 1)\quad ,
$
$
[\, J_{+} , J_{-}\, ] = - 2 J_{3} \quad , \quad [\, J_{3}  , J_{\pm}\, ] = \pm
J_{\pm} \quad .$ The Casimir operator ${\cal C}$ is
${{\cal C}^{2} \equiv {1\over {4}} + J_{3}^{2} - {1\over{2}} (
J_{+} J{-} + J_{-} J_{+} )}$ ${= {1\over{4}} ( A^{\dagger} A -
B^{\dagger} B)^{2}} $.

We also observe that
$
[\, H_{0} , H_{I}\, ] = 0
$.
The time evolution of the vacuum
$
|0> \equiv | n_{A} = 0 , n_{B} = 0 > ~,~
A |0> = 0 = B |0>~
$,
is controlled by $H_{I}$
$$
\eqalign{
|0(t)> = \exp{ \left ( - i t {H \over{\hbar}} \right )} |0> =
\exp{ \left ( - i t {H_{I} \over{\hbar}} \right )} |0>   \cr
= {1\over{\cosh{(\Gamma t)}}} \exp{
\left ( \tanh {(\Gamma t)} A^{\dagger} B^{\dagger} \right )}|0> \quad ,}
\eqno{(2.8a)}
$$
$$
<0(t) | 0(t)> = 1~ \quad \forall t~,
\eqno{(2.8b)}
$$
$$
\lim_{t\to \infty} <0(t) | 0> \, \propto \lim_{t\to \infty}
\exp{( - t  \Gamma )} = 0 \quad .
\eqno{(2.9)}
$$
Time evolution transformations for creation and annihilation operators are
$$
\alpha \mapsto \alpha(t) =
{\rm e}^{- i {t\over{\hbar}} H_{I}}
\alpha ~{\rm e}^{i {t\over{\hbar}} H_{I}} =
\alpha \cosh{(\Gamma t)} - \alpha^{\dagger} \sinh{(\Gamma t)} ~,
 \eqno{(2.10a)}
$$
$$
\beta \mapsto \beta(t) =
{\rm e}^{- i {t\over{\hbar}} H_{I}}
\beta ~{\rm e}^{i {t\over{\hbar}} H_{I}} =
\beta \cosh{(\Gamma t)} + \beta^{\dagger} \sinh{(\Gamma t)}
\quad   \eqno{(2.10b)}
$$
and h.c., and the corresponding ones for $A(t)$, $B(t)$ and h.c..
We note that eqs. 2.10 are canonical transformations preserving the
ccr eq. 2.5b.
Eq.2.9 expresses the
instability (decay) of the vacuum under the evolution
operator $\exp{ \left ( - i t {H_{I} \over{\hbar}} \right )}$. This means that
the QM framework is not suitable for the canonical quantization
of the dho. In other words time evolution leads out of the Hilbert space of
the states
and
in ref. 3 it has been shown that the proper way to perform
the canonical quantization of the dho is to work in the framework
of QFT. In fact
for many degrees of freedom the time evolution operator ${\cal U}(t)$
and the vacuum are formally (at finite volume) given by
$$
\eqalign{
{\cal U}(t) =
\prod_{\kappa}{\exp\Bigl(-{\Gamma_{\kappa} t \over{ 2}}\bigl(
{\alpha}_{\kappa}^2 -
{\alpha}_{\kappa}^{\dagger 2}\bigr)
\Bigr)
\exp\Bigl({\Gamma_{\kappa} t \over{ 2}}\bigl(
{\beta}_{\kappa}^2 -
{\beta}_{\kappa}^{\dagger 2}\bigr)
\Bigr)}
\cr
=\prod_{\kappa}{\exp{\Bigl ( \Gamma_{\kappa} t \bigl ( A_{\kappa}^{\dagger}
B_{\kappa}^{\dagger} - A_{\kappa} B_{\kappa} \bigr ) \Bigr )}},}
\eqno{(2.11)}
$$
$$
|0(t)> = \prod_{\kappa} {1\over{\cosh{(\Gamma_{\kappa} t)}}} \exp{
\left ( \tanh {(\Gamma_{\kappa} t)} A_{\kappa}^{\dagger}
B_{\kappa}^{\dagger} \right )} |0> \quad ,
\eqno{(2.12)}
$$
\noindent with
$<0(t) | 0(t)> = 1~, \quad \forall t $~.
Using the continuous limit relation $
\sum_{\kappa} \mapsto {V\over{(2 \pi)^{3}}} \int \! d^{3}{\kappa}$,
in the
infinite-volume limit we have (for $\int \!
d^{3} \kappa ~
\Gamma_{\kappa}$ finite and positive)
$$
{<0(t) | 0> \rightarrow 0~~ {\rm as}~~ V\rightarrow \infty }
{}~~~\forall~  t~  , \eqno{(2.13)}
$$
and in general,
$
{<0(t) | 0(t') > \rightarrow 0~~ {\rm as}~~ V\rightarrow \infty}
{}~~~\forall~t~ and~ t'~ ,~~~ t' \neq t.
$
At each time {\twelveit t} a representation
$\{ |0(t)> \}$ of the ccr is defined and turns out to be {\twelveit ui}
to any other
representation $\{ |0(t')>~,~~\forall t'\neq t \}$ in the infinite volume
limit. In such a way the quantum dho evolves in time through {\twelveit ui}
representations of ccr ({\twelveit tunneling}).
We remark that  $| 0(t)>$ is a two-mode time dependent
Glauber coherent state$^{22,23}$.

We thus see that the Bogolubov transformations,
corresponding to eqs.2.10,
can be implemented for every $\kappa$ as inner automorphism for the
algebra  ${su(1,1)}_{\kappa}$. At each time  {\twelveit t}
we have a copy
$\{ A_{\kappa}(t) , A_{\kappa}^{\dagger}(t) , B_{\kappa}(t) ,
B_{\kappa}^{\dagger}(t) \, ; \, | 0(t) >\, |\, \forall {\kappa} \}$
of the
original algebra induced by the time evolution operator which
can thus be thought of as a generator
of the group of automorphisms of ${\bigoplus_{\kappa}
su(1,1)_{\kappa}}$ parameterized by time  {\twelveit t} (we have a
realization of the operator algebra at
each time t,
which can be implemented by Gel'fand-Naimark-Segal construction in the
C*-algebra formalism$^{10}$).
Notice that the various copies
become unitarily inequivalent in the infinite-volume limit, as
shown by eqs.2.13: the space of the states splits
into {\twelveit ui} representations of the ccr each one labelled by time
parameter  {\twelveit t}.
As usual one works at finite volume and only at the end
of the computations the limit $V \to \infty$ is performed.

Finally, in refs. 2 and 3 it has been shown that the
representation $\{|0(t)>\}$ is equivalent to the TFD representation
$\{|0(\beta(t)>\}$, thus recognizing the relation between the dho
states and the finite temperature states.
In particular, one may introduce the { \twelveit free energy}
functional for the $A$-modes
$$
{\cal F}_{A} \equiv <0(t)| \Bigl (  H_{A} - {1\over{\beta}} S_{A}
\Bigr ) |0(t)> \quad , \eqno{(2.14)}
$$
where $H_{A}$ is the part of $H_{0}$ relative to  $A$-
modes only,
namely $H_{A} = \sum_{\kappa} \hbar \Omega_{\kappa}
A_{\kappa}^{\dagger} A_{\kappa}$, and the {\twelveit entropy} $S_{A}$
is given by
$$
 S_{A} \equiv - \sum_{\kappa} \Bigl \{ A_{\kappa}^{\dagger} A_{\kappa}
\ln \sinh^{2} \bigl ( \Gamma_{\kappa} t \bigr ) - A_{\kappa}
A_{\kappa}^{\dagger} \ln \cosh^{2} \bigl ( \Gamma_{\kappa} t \bigr ) \Bigr \}
\quad .  \eqno{(2.15)}
$$
One then considers
the stability condition
${{\partial {\cal F}_{A}}\over{\partial \vartheta_{\kappa}}} = 0 \quad
 \forall \kappa \quad ,\vartheta_{\kappa} \equiv \Gamma_{\kappa} t$~
to be satisfied in each representation,
and using the definition $E_{\kappa} \equiv \hbar \Omega_{\kappa}$, one finds
$$
{\cal N}_{A_{\kappa}}(t) = \sinh^{2} \bigl ( \Gamma_{\kappa} t \bigr ) =
{1\over{{\rm e}^{\beta (t) E_{\kappa}} - 1}} \quad , \eqno{(2.16)}
$$
\noindent which is the Bose distribution for $A_{\kappa}$ at time
{\twelveit t}.
$\{ |0(t)> \}$ is  thus recognized to be a representation of
the ccr's at finite temperature, equivalent
to the TFD representation $\{ |0(\beta )> \}$~$^{11,16,17}$.
Furthermore, use of eq.2.15 shows that
$$
{{\partial}\over{\partial t}} |0(t)> =  - \left ( {1\over{2}}
{{\partial {\cal S}}\over{\partial t}} \right ) |0(t)> \quad . \eqno{(2.17)}
$$
We thus see
that $i \left ( {1\over{2}} \hbar {{\partial
{\cal S}}\over{\partial t}} \right )$ is the
generator of time translations,
namely
time evolution
is controlled by the entropy variations$^{24}$.
It appears to
us remarkable that the same
dynamical variable
${\cal S}$
whose expectation value is formally the entropy also
controls time evolution: Damping
(or, more generally, dissipation) implies indeed the choice of a privileged
direction in time evolution ({\twelveit
arrow of time}) with a consequent breaking of
time-reversal invariance.
We may also show that
$
d {\cal F}_{A} = d E_{A} - {1\over{\beta}} d {\cal S}_{A}=0~,
$ which
expresses the
first principle of thermodynamics for a system coupled with environment
at constant temperature and in absence of mechanical work.
We may define as usual heat as ${dQ={1\over{\beta}} dS}$
and see that the change in time $d {\cal N}_{A}$ of particles condensed in the
vacuum turns out into heat dissipation $dQ$.
\bigskip
\noindent{\twelvebf 3. Quantum dissipation and quantum noise }
\bigskip
Let us now ask the following question:
Does the introduction of an ``extra coordinate'' make any sense in the
context of conventional QM?
To answer to such a question we
consider the special case
of zero mechanical resistance. One then begins with the
Hamiltonian for an isolated particle and the corresponding density
matrix equation
$$
H=-(\hbar^2/2m )(\partial/\partial Q)^2 +V(Q). \eqno(3.1)
$$
$$
i\hbar (\partial \rho /\partial t)=[H,\rho ], \eqno(3.2)
$$
which indeed requires two coordinates (say $Q_+$ and $Q_-$). In the
coordinate representation, we have$^{4}$
$$
i\hbar (\partial/\partial t)<Q_+|\rho (t)|Q_->=
$$
$$
\{
-(\hbar^2/2m)[(\partial/\partial Q_+)^2-(\partial/\partial Q_-)^2]
+[V(Q_+)-V(Q_-)]
\}<Q_+|\rho (t)|Q_->. \eqno(3.3)
$$
In terms of the coordinates $x$ and $y$, it is $
Q_{\pm}=x\pm (1/2)y$,~
and the density matrix function
$
W(x,y,t)=<x+(1/2)y|\rho (t)|x-(1/2)y>
$.
{}From eq.3.3 the Hamiltonian now reads
$
{\cal H}_o=(p_xp_y/m)+V(x+(1/2)y)-V(x-(1/2)y),
$ with
$
p_x=-i\hbar(\partial /\partial x),
\ \ p_y=-i\hbar(\partial /\partial y),
$ which,
of course, may be constructed from the
``Lagrangian''
$$
{\cal L}_{0}(\dot{x},\dot{y},x,y)=m \dot{x}\dot{y}
-V(x+(1/2)y)+V(x-(1/2)y), \eqno(3.4)
$$
We have then the justification for introducing eq.2.2
at least for the case $\gamma=0$. Notice indeed that for $
V(x\pm (1/2)y)=(1/2)k(x\pm (1/2)y)$ eq.3.4 gives eq.2.2
for the case $\gamma =0$.
We also notice that ${\cal H}_{o}$ and $\cal H_{I}$ in eq.2.7
are the free Hamiltonian and the generator of Bogolubov
transformations,respectively, in TFD$^{11,16,17}$. Our present discussion
thus includes the doubling of degrees of freedom in finite temperature
QFT.

Next,
our task is to explore the manner in which the Lagrangian model for
quantum dissipation of refs. 1-5,13 arises
from the formulation of the quantum Brownian
motion problem as described by Schwinger$^{8}$
and by Feynman and Vernon$^{9}$.

Let us suppose that the particle interacts with a thermal bath
at temperature $T$. The interaction Hamiltonian between the bath and
the particle is taken as
$
H_{int}=-fQ,
$
where $Q$ is the particle coordinate and $f$ is the random force on
the particle due to the bath.
In the Feynman-Vernon formalism the effective
action for the particle has the form
$$
{\cal A}[x,y]=\int_{t_i}^{t_f}dt{\cal L}_o(\dot{x},\dot{y},x,y)
+{\cal I}[x,y], \eqno(3.5)
$$
where ${\cal L}_o$ is defined in eq.3.4 and
$$
e^{(i/\hbar){\cal I}[x,y]}=
<(e^{(-i/\hbar)\int_{t_i}^{t_f}f(t)Q_-(t)dt)})_-
(e^{(i/\hbar)\int_{t_i}^{t_f}f(t)Q_+(t)dt)})_+>. \eqno(3.6)
$$
In eq.3.6 the average is with respect to the thermal bath;
``$(.)_{+}$'' denotes time ordering and ``$(.)_{-}$''
denotes anti-time ordering.
If the interaction between the bath and the coordinate $Q$
were turned off, then the operator $f$ of the bath
would develop in time according to
$f(t)=e^{iH_R t/\hbar}fe^{-iH_R t/\hbar }$ where $H_R$ is the Hamiltonian
of the isolated bath (decoupled from the coordinate $Q$).
$f(t)$ is the force operator of the bath to be used in eq.3.6.
Assuming that the particle makes contact with the bath at the
initial time $t_i$, the reduced density matrix function is
at a final time
$$
W(x_f,y_f,t_f)=\int_{-\infty}^{\infty }dx_i\int_{-\infty}^{\infty }dy_i
K(x_f,y_f,t_f;x_i,y_i,t_i)W(x_i,y_i,t_i), \eqno(3.7)
$$
$$
K(x_f,y_f,t_f;x_i,y_i,t_i)=
\int_{x(t_i)=x_i}^{x(t_f)=x_f}{\cal D}x(t)
\int_{y(t_i)=y_i}^{y(t_f)=y_f}{\cal D}y(t)
e^{(i/\hbar){\cal A}[x,y]}. \eqno(3.8)
$$

The correlation function for the random force on the particle
is given by
$
G(t-s)=(i/\hbar )<f(t)f(s)>.
$
The retarded and advanced Greens functions are defined by
$
G_{ret}(t-s)=\theta (t-s)[G(t-s)-G(s-t)],
$
and
$
G_{adv}(t-s)=\theta (s-t)[G(s-t)-G(t-s)]~.
$
The mechanical resistance is defined $
R=lim_{\omega \rightarrow 0}{\cal R}e Z(\omega +i0^+),
$ ~with
the mechanical impedance
$Z(\zeta )$ (analytic in the upper half complex frequency
plane ${\cal I}m
{}~\zeta >0$) determined by the retarded Greens function
$
-i\zeta Z(\zeta )=\int_0^\infty dt G_{ret}(t)e^{i\zeta t}.
$
The time domain quantum noise in the fluctuating random force is
$
N(t-s)=(1/2)<f(t)f(s)+f(s)f(t)>.
$

The time ordered and anti-time ordered Greens functions
describe both the retarded and advanced Greens functions as well
as the quantum noise,
$$
G_\pm (t-s)=\pm (1/2)[G_{ret}(t-s)+G_{adv}(t-s)]
+(i/\hbar )N(t-s). \eqno(3.9)
$$

The interaction between the bath and the particle is evaluated by following
Feynman and Vernon and we find$^{4}$ for the real and the imaginary
part of the action
$$
{\cal R}e{\cal A}[x,y]=\int_{t_i}^{t_f}dt{\cal L}, \eqno(3.10a)
$$
$$
{\cal L}=m \dot{x}\dot{y}-[V(x+(1/2)y)-V(x-(1/2)y)]
+(1/2)[xF_y^{ret}+yF_x^{adv}], \eqno(3.10b)
$$
$$
{\cal I}m{\cal A}[x,y]=
(1/2\hbar )\int_{t_i}^{t_f}\int_{t_i}^{t_f}dtdsN(t-s)y(t)y(s),
\eqno(3.10c)
$$
respectively,
where the retarded force on $y$ and the advanced force on $x$ are
defined as
$
F_y^{ret}(t)=\int_{t_i}^{t_f}dsG_{ret}(t-s)y(s),~~
F_x^{adv}(t)=\int_{t_i}^{t_f}dsG_{adv}(t-s)x(s). $

Eqs.3.10 are {\twelveit rigorously exact} for linear passive damping
due to the bath when the path integral eq.3.8 is employed for the
time development of the density matrix.

We therefore conclude that
the lagrangian eq.2.2 can be viewed as the approximation
to eq.3.10b with $F_y^{ret}=\gamma\dot{y}$ and $F_x^{adv}=-\gamma\dot{x}$.

We also observe that at the classical level
the ``extra'' coordinate $y$, is usually constrained to vanish.
(Note that $y(t)=0$ is a true solution to
eqs.2.2 so that the constraint is {\twelveit
not} in violation of the equations
of motion.)
{}From eqs.3.10 we thus also conclude that
the classical constraint $y=0$ occurs because nonzero $y$ yields an
``unlikely process'' in view of the large imaginary part of the action
(in the classical ``$\hbar \rightarrow 0$''limit) implicit in eq.3.10c.
On the contrary, at quantum level nonzero $y$ allows quantum
noise effects arising from the imaginary part of the action.
\bigskip
\noindent {\twelvebf 4. Quantum dissipation
and the q-deformation of the WH algebra}
\bigskip
Quantum deformations$^{20,25}$
of Lie algebras are well studied mathematical
structures and therefore their properties need not to be presented
again in this paper; we only recall that they are
deformations of the enveloping algebras of Lie algebras and have
Hopf algebra structure. In particular the q-WH algebra has the
properties of graded Hopf algebra$^{26}$.
In this section our task is to establish a formal relation
between the q-WH algebra and the dho hamiltonian.
To this aim we
introduce the realization of q-WH algebra in terms
of finite difference operators$^{27}$.
Since we want to preserve the analytic properties of
Lie algebra in the deformation procedure, we work in
the Fock-Bargmann representation (FBR)
in QM$^{23}$.
In the FBR the operators
$
N \to z {d\over dz}~,~~
a^\dagger \to z~,~~ a \to {d\over dz}~,~~
z~ \in {\CcC}~~,
$
provide a realization of the WH algebra
$
[ a, a^\dagger ] = \IiI~,~~ [ N, a ] = - a~,~~ [ N,
a^\dagger ] = a^\dagger~~~.
$
The Hilbert space ~$\cal F$ is identified with the space
of the entire
analytic functions and has well defined inner product.

We consider the finite difference operator ${\cal D}_q$
defined by:
$
{\cal D}_q ~f(z)={{f(q z) - f(z)}\over {(q-1) ~z}}=$
${{q^{z {d\over {dz}}} - 1}\over{(q-1)~ z }}f(z)~, $
with ~$f \in {\cal F}~, q = e^\zeta , ~\zeta \in {\CcC}$.
${\cal D}_q$ is called the
q-derivative operator and, for $q \to 1$ (i.e.
$\zeta \to 0$), it reduces to the standard derivative.
Next, we introduce the following operators
in the space ${\cal F}$:
$
N \to z {d\over dz} ~, ~{\hat a}_q \to z~,
{}~a_q \to {\cal D}_q ~, $
where  ~${\hat a}_q = {\hat a}_{q=1} =
a^\dagger$ and ~${\lim_{q\to1} a_q =a}$.
The q-WH algebra, in terms of the operators
$\{ a_q, {\hat a}_q, N\equiv N_q ;
 ~q \in \CcC \}$,  is then realized by the relations:
$$
[ N, a_q ] = - a_q~~,~~[ N, {\hat a}_q ] =
{}~~{\hat a}_q \; ,\;
[ a_q, {\hat a}_q ]~\equiv~a_q {\hat a}_q~-~{\hat a}_q a_q~=
{}~q^N  . \eqno{(4.1)}
$$
By introducing ~~${\bar a}_q ~\equiv ~
{\hat a}_q q^{-N/2}$~, it assumes the more familiar form$^{20}$
$
[ N , a_q ]=- a_q$,$[ N , {\bar a}_q ] ={\bar a}_q \;
,  \;a_q {\bar a}_q - q^{-{1\over 2}} {\bar a}_q a_q=q^{{1\over
2}N}\; .
$
We can show that the commutator  $[ a_q, {\hat a}_q]$
acts in ${\cal F}$ as follows$^{27}$

$$ [ a_q, {\hat a}_q ]f(z)~=~q^{z{d\over dz}}f(z)~=~f(qz)~.
\eqno{(4.2)}$$

The q-deformation of the WH algebra is thus strictly related with
the finite difference operator ~${\cal D}_q$ ($q \not= 1$).
This leads us to conjecture that
the q-deformation of the operator algebra
should arise whenever we are
in the presence of
lattice or discrete
structure$^{27}$.
In the following we show that this indeed happens in the case of damping
where the finite life-time
$\tau = {1 \over{\Gamma}}$
acts as time-unit for the system.

We now express the time evolution
generator of dho in terms  of the commutator
$ [ a_q, {\hat a}_q ]$ of the q-WH algebra.
In the FBR Hilbert space $~{\cal F}$ we introduce
$$
 \tilde a \equiv {1\over {\sqrt{2\hbar\Omega}}} \bigl({ p_z\over{\sqrt{m}}}
- i{\sqrt{m} \Omega} z
\bigr) , ~~ { \tilde a}^\dagger \equiv {1\over {\sqrt{2\hbar\Omega }}}
\bigl({ p_z\over{\sqrt{m}}} +  i{\sqrt{m}\Omega} z
\bigr), ~~~
[\tilde a, {\tilde a}^\dagger ] = \IiI   \eqno{(4.3)}
$$
where $z \in {\CcC}$, $p_z = - i \hbar {d\over {dz}}$ and $[ z, p_z]
= i \hbar$.
The conjugation of $\tilde a$ and ${\tilde a}^\dagger$ is as usual well
defined with respect to the inner product defined in $~{\cal F}$$^{23}$.

We note that, by setting $ Re(z)=x$, ${\tilde a} \to \alpha$ and
${\tilde a}^{\dagger} \to {\alpha}^{\dagger}$ in the limit
$Im(z) \to 0$, where $\alpha$ and ${\alpha}^{\dagger}$ are
the annihilation
and creation operators introduced in eqs.2.5.
By putting $q={\rm e}^{\theta}$ with $\theta$ real
we can check that the operator
$$
[ a_q, {\hat a}_q]  ~=~
  \exp\Bigl(\theta z {d\over dz}\Bigr) ~=~
{1\over{\sqrt q}} \exp\Bigl(-{\theta\over{ 2}}\bigl({\tilde a}^2 -
{\tilde a}^{\dagger 2}\bigr)
\Bigr)
\equiv {1\over{\sqrt {q}}}{\hat {\cal S}}(\theta) ,~~
\eqno{(4.4)}
$$
generates the Bogolubov transformations:
$$
\tilde a(\theta) = {\hat{\cal S}}(\theta)
\tilde a {\hat {\cal S}}(\theta)^{-1} =
 \tilde a \cosh{\theta} - \tilde a^{\dagger} \sinh{
\theta}
$$
$$
\mapsto \alpha(\theta)  =
 \alpha \cosh{\theta} - {\alpha}^{\dagger} \sinh{
\theta}
{}~~~~~ as~~ Im(z) \to 0~.
\eqno{(4.5)}
$$
and h.c..
Note that
the right hand side of eq. 4.4 is an $SU(1,1)$ group element.
In fact ${1\over  2}{\tilde a} ^2 = K_{-}$,
${1\over  2}{\tilde a}^{\dagger 2} = K_{+}$,
${1\over 2}(\tilde a^\dagger
\tilde a + {1\over 2})
= K_{3}$,
close the $su(1,1)$ algebra.
We remark that the transformation eq. 4.5,
which is a canonical transformation,
has been shown$^{21}$ to relate the Weyl representations of the ccr in
QM$^{23}$ and thus the deformation parameter $q={\rm e}^{\theta}$
acts as a label for the Weyl representations.

Next, we introduce the q-WH algebra operators $b_{q'}$ and ${\hat b}_{q'}$
corresponding to the doubled degree of freedom $\beta$ introduced
in sec. 2;
by introducing the operators
$$
 \tilde b \equiv {1\over {\sqrt{2\hbar\Omega}}} \bigl({ p_{\zeta}\over{
\sqrt{m}}}
- i{\sqrt{m} \Omega} {\zeta}
\bigr) , ~~ { \tilde b}^\dagger \equiv {1\over {\sqrt{2\hbar\Omega }}}
\bigl({ p_{\zeta}\over{\sqrt{m}}} +  i{\sqrt{m}\Omega} {\zeta}
\bigr), ~~~
[\tilde b, {\tilde b}^\dagger ] = \IiI   \eqno{(4.6)}
$$
with $\zeta \in {\CcC}$, $Re(\zeta) \equiv y$,
$p_{\zeta} = - i \hbar {d\over {d \zeta}}$ and
$[ \zeta, p_{\zeta}] = i \hbar$, we proceed as in the above case of
$\tilde a$ operator and obtain

$$
[ a_{q}, {\hat a}_{q}][ b_{q'}, {\hat b}_{q'}] =
\exp \Bigl(-{{\theta} \over 2}
\bigl[\bigl({{\tilde a}} ^2 -
{{\tilde a}}^{\dagger 2} \bigr) - \bigl({{\tilde b}}^{2} -
{{\tilde b}}^{\dagger 2} \bigr)\bigr] \Bigr),
\eqno{(4.7)}
$$
with $q'={q}^{-1}$. We finally see that,
provided
we set $q={\rm e}^{\theta}$, with ${\theta} \equiv {\Gamma} t$,
the operator in eq.4.7
acts as the time evolution operator
${\cal U}(t)$
(cf. eq.2.11) in the limits $Im(z) \to 0$ and $Im(\zeta) \to 0$:
$$
[ a_{q}, {\hat a}_{q}][ b_{q'}, {\hat b}_{q'}]
\mapsto
\exp \Bigl(-{{\Gamma t} \over 2}
\bigl[\bigl({{\alpha}} ^2 -
{{\alpha}}^{\dagger 2} \bigr) - \bigl({{\beta}}^2 -
{{\beta}}^{\dagger 2} \bigr)\bigr] \Bigr)=
{\cal U}(t)
\eqno{(4.8)}
$$
\indent Eq.4.8 is the wanted expression of
 the time-evolution generator of the dho in terms of the q-WH
algebra operator $[ a_{q}, {\hat a}_{q}][ b_{q'}, {\hat b}_{q'}]$.

Our discussion can be generalized to the case of many degrees of
freedom by observing that eq.4.8
holds for each couple of modes (${\alpha}_{\kappa},{\beta}_{\kappa}$)
and formally we have
$$
\prod_{\kappa}{[ a_{\kappa,q}, {\hat a}_{\kappa,q}][ b_{\kappa,q'},
{\hat b}_{\kappa,q'}]} ~\to~
\exp\bigl(-{i\over{\hbar}} H_{I}t \bigr)={\cal U}(t)~~, ~~~
as~Im\{z,\zeta \}  \to 0~~,~
\eqno{(4.9)}
$$
where $q={\rm e}^{\theta_\kappa}$,
${\theta_\kappa} \equiv {\Gamma_\kappa}t$,  $q'={q}^{-1}$,
$Im\{z,\zeta\}$ denotes the imaginary part of FBR $z$- and
$\zeta$-variables associated
to each ${\alpha}_{\kappa}$ and ${\beta}_{\kappa}$ mode and ${\cal U}(t)$
is given by eq.2.11.
Notice that for simplicity of notation here $q$ denotes $q_{\kappa}$.
We note that, through its time
dependence, the deformation
parameter $q(t)$ labels the {\twelveit ui} representations $\{|O(t)>\}$.

We also observe that eq.4.9 leads to
representation of the generator
of the thermal Bogolubov transformation in terms of the operators
$\prod_{\kappa}{[ a_{\kappa,q}, {\hat a}_{\kappa,q}]
[ b_{\kappa,q'}, {\hat b}_{\kappa,q'}]}$:
indeed, by resorting to the discussion of sec. 2 where the representation
$\{|O(t)>\}$ for the dho has been
shown to be the same as the
TFD representation $\{|O(\beta (t))>\}$ (see refs. 2,3),
we also recognize the strict relation between q-WH algebra and
finite temperature QFT.
\bigskip
\line{{\twelvebf 5. Comments and conclusion}\hfill}
\bigskip
The dho total Hamiltonian is invariant under the transformations
generated by ${J_{2} = \bigoplus_{\kappa} J_{2}^{(\kappa )}}$. The
vacuum however is not invariant under $J_{2}$ (see eq.2.8a) in the infinite
volume limit. Moreover, at each time t, the representation $\{|0(t)>\}$~
may be
characterized by the expectation value in the state $|0(t)>$  of,
{\twelveit e.g.},
${J_{3}^{(\kappa )} - {1\over{2}}}$: thus the total number of
particles~ $n_{A} +n_{B} = 2n$~ can be taken as an order parameter. Therefore,
at each time t the symmetry under $J_{2}$ transformations is spontaneously
broken. On the other hand, ${\cal H}_{I}$ is proportional to $J_{2}$. Thus, in
addition to breakdown of time-reversal (discrete) symmetry, already mentioned
in sec. 2, we also have spontaneous breakdown of time
translation (continuous) symmetry.
In other words we have described dissipation (i.e. energy
non-conservation), as an effect of
breakdown of time translation and time-reversal symmetry. It is an interesting
question asking which is the zero-frequency mode, playing the r\^ole of
the Goldstone mode,
related with the breakdown of continuous time translation symmetry: we observe
that since~ $n_{A} - n_{B}$ is constant in time, the condensation (annihilation
and/or creation) of AB-pairs does not contributes to the vacuum energy so that
AB-pair may play the r\^ole of a zero-frequency mode.

\indent   From the point of view of boson condensation, time evolution in the
presence of damping may be thought of as a sort of continuous transition
among different phases, each phase corresponding, at time  {\twelveit t},
to the
coherent state representation $\{|0(t)>\}$.
The damped oscillator thus provides an archetype of
system  undergoing continuous phase transition.

In the discussion presented above a crucial r\^ole is
played by the existence of infinitely many {\twelveit ui}
representations of the
ccr in QFT. In ref. 21 the q-WH algebra has been discussed in
relation with the von Neumann theorem in QM and it has been
shown on a general ground that
the q-deformation parameter acts as a label for the
{\twelveit Weyl systems} in QM and for the {\twelveit ui}
representations in QFT; the
mapping between {\twelveit different}
 ($i.e.$ labelled by different values of $q$)
representations (or Weyl systems) being
performed by the Bogolubov transformations (at finite volume).
Damped
harmonic oscillator and
finite temperature systems are explicit examples clarifying
the physical meaning of such a labelling
(further examples are provided by unstable particles in QFT$^{24}$,
by
quantization of the matter field in curved space-time$^{28}$, by
theories with spontaneous breakdown of symmetry where  different
values of the order parameter are associated to
different {\twelveit ui} representations (different {\twelveit phases})).
In the case of damping,
as well as in the
case of time-dependent temperature,
the system time-evolution is represented as
{\twelveit tunneling} through {\twelveit
ui} representations: the non-unitary
character of time-evolution ({\twelveit arrow of time}) is thus expressed
by the non-unitary equivalence of the representations in the infinite
volume limit. It is remarkable that at the algebraic level this is
made possible through the
q-deformation mechanism which
organizes the representations in an {\twelveit ordered set} by means of
the labelling.

In ref. 27 it has been also shown that
the commutator $[ a_q, {\hat a}_q]$ acts as squeezing
generator (indeed the operator ${\hat S}(\theta)$ in eq.4.4 acts like
the squeezing generator with respect to ${\tilde a}$ and ${\tilde a}^\dagger$
operators), a result
which confirms
the relation between dissipation and squeezed coherent states
exhibited in ref. 1. In turn, q-groups have been also shown$^{29}$
to be
the  natural candidates to study squeezed coherent state.

Finally, we mention that dissipative systems have been discussed in the
framework of stochastic quantization and coherent states in ref. 30.

We are glad to acknowledge the
Directors of The Workshop, E.C.G. Sudarshan
and V.G.Vaccaro, and the Director of the INFN Eloisatron Project,
A. Zichichi, for the kind invitation and hospitality at The Ettore
Majorana Centre.
\phantom{xxxxx}
\bigskip
{\twelvebf References}
\bigskip
\item{1.}	E. Celeghini, M. Rasetti, M. Tarlini and
	G. Vitiello, { \twelveit Mod.
	Phys. Lett.} {\twelvebf B 3 } (1989) 1213

\item{2.}	E.Celeghini, M.Rasetti and G.Vitiello in {\twelveit
        Thermal Field
	Theories andTheir Applications}, H.Ezawa, T.Aritmitsu
	and Y.Hashimoto Eds., Elsevier, Amsterdam 1991, p. 189

\item{3.}   E. Celeghini, M. Rasetti and G. Vitiello, {\twelveit
        Annals of Phys.
	(N.Y.)} {\twelvebf 215} (1992) 156

\item{4.}	Y.N.Srivastava, G.Vitiello and A.Widom, {\twelveit
        Quantum Dissipation
	and quantum noise}, {\twelveit Annals of Phys. (N.Y.)}, in press

\item{5.} A.Iorio and G.Vitiello, {\twelveit Quantization of damped harmonic
	oscillator, thermal field theory and q-groups},
        in {\twelveit Proceedings of The Third
        International Workshop on Thermal Field Theories},
        Banff (Canada) 1993; {\twelveit Quantum dissipation and quantum
        groups}, Salerno preprint 1994

\item{6.} E.C.G.Sudarshan, C.B.Chiu and G.Bhamathi, {\twelveit Unstable
        systems in Generalized Quantum Theory}, DOE-40757-023
        CPP-93-23,1993

\item{7.}  C.B.Chiu, E.C.G.Sudarshan and G.Bhamathi, {\twelveit Phys. Rev.}
        {\twelvebf D45} (1992) 884

\item  {\phantom{7}} E.C.G.Sudarshan and C.B.Chiu, {\twelveit Phys. Rev.}
        {\twelvebf D47} (1993) 2602

\item{8.} J. Schwinger, {\twelveit
       J. Math. Phys.} {\twelvebf 2} (1961), 407

\item{9.} R.P. Feynman and F.L. Vernon, {\twelveit Annals of Phys. (N.Y.)}
{\twelvebf 24} (1963) 118

\item{10.} O.Bratteli and D.W.Robinson, {\twelveit
    Operator Algebras and Quantum
Statistical Mechanics} (Springer, Berlin, 1979)

\item{11.} H.Umezawa, {\twelveit Advanced field theory: micro,
macro and thermal concepts} (American Institute of Physics, N.Y. 1993)

\item{12.} H. Bateman, {\twelveit Phys. Rev.} {\twelvebf 38} (1931), 815

\item{13.} H.Feshbach and Y.Tikochinsky, { \twelveit
        Transact. N.Y. Acad. Sci.}
	{\twelvebf 38} (Ser. II) (1977) 44

\item{14.} G.Lindblad and B.Nagel, { \twelveit
       Ann. Inst. H. Poincar\'e} {\twelvebf
	XIII A} (1970) 27

\item{15.} H.P.Yuen, {\twelveit Phys. Rev.} {\twelvebf A13} (1976) 2226

\item{16.}  Y. Takahashi and H. Umezawa, { \twelveit Collective Phenomena}
         {\twelvebf 2}
	(1975) 55

\item{17.} H. Umezawa, H. Matsumoto and M. Tachiki, {\twelveit
	Thermo Field Dynamics
        and Condensed States}, North-Holland Publ.
        Co., Amsterdam 1982

\item{18.} P.Shanta, S.Chaturvedi, V.Srinivasan and F.Mancini,
         {\twelveit Mod.Phys.Lett.} {\twelvebf A8} (1993) 1999

\item{19.} Y. Tsue, A. Kuriyama and M. Yamamura, {\twelveit Thermal
         effects and dissipation
         in SU(1,1),}
         Kochi University
         preprint 1993

\item{20.}   L.C.Biedenharn, {\twelveit J.Phys.}  {\twelvebf A22} (1989) L873

\item	{\phantom{20}} A.J.Macfarlane, {\twelveit J. Phys.} {\twelvebf A22}
      	(1989) 4581

\item{21.}   A.Iorio and G.Vitiello, {\twelveit Mod. Phys. Lett.}
        {\twelvebf B 8},
         (1994) 269

\item{22.} J.R.Klauder and E.C.Sudarshan, {\twelveit Fundamentals of Quantum
	Optics}, Benjamin, New York 1968

\item{23.} A.Perelomov, {\twelveit Generalized Coherent States and Their
        Applications}, Springer-Verlag,
         Berlin, Heidelberg 1986

\item{24.} S. De Filippo and G. Vitiello, { \twelveit Lett. Nuovo Cimento}
      {\twelvebf 19}
	(1977) 92

\item{25.} Drinfeld V.G., Proc. ICM Berkeley, CA; A.M. Gleason, ed,; AMS,
	   Providence, R.I., 1986, page 798.

\item   {\phantom{25}} M.Jimbo, {\twelveit Int. J. of Mod. Phys.} {
       \twelvebf A4} (1989) 3759.

\item	{\phantom{25}} Yu.I.Manin, {\twelveit
        Quantum groups and Non-Commutative Geometry},
        CRM, Montreal, 1988.

\item{26.} 	E.Celeghini, T.D.Palev and M.Tarlini,
       {\twelveit Mod. Phys.Lett.}
        {\twelvebf B5 } (1991) 187

\item 	{\phantom{26}} P.P.Kulish and N.Yu.Reshetikin,
     {\twelveit Lett. Math.
	Phys.} {\twelvebf 18} (1989) 143

\item{27.}  E.Celeghini, S.De Martino, S.De Siena, M.Rasetti and
        G. Vitiello, {\twelveit
       Mod. Phys. Lett.} {\twelvebf B 7} (1993) 1321;
        {\twelveit Quantum Groups, Coherent States, Squeezing
	and Lattice Quantum Mechanics}, preprint 1993

\item{28.} M. Martellini, P. Sodano and G. Vitiello, { \twelveit
         Nuovo Cimento}
         {\twelvebf
	48 A} (1978) 341

\item{29.} E.Celeghini, M.Rasetti and G.Vitiello, {\twelveit Phys. Rev.
	Lett.} {\twelvebf 66} (1991), 2056

\item{30.} S.De Martino, S.De Siena, F.Illuminati and G.Vitiello,
        {\twelveit Diffusion processes and coherent states}, {\twelveit
        Mod. Phys. Lett}{\twelvebf B} in print

\vfill\eject
\bye